\documentclass[sigconf]{acmart}
\AtBeginDocument{%
  }

\copyrightyear{2025}
\acmYear{2025}
\setcopyright{acmlicensed}
\acmConference[FSE Companion '25]{33rd ACM International Conference on the Foundations of Software Engineering}{June 23--28, 2025}{Trondheim, Norway}
\acmBooktitle{33rd ACM International Conference on the Foundations of Software Engineering (FSE Companion '25), June 23--28, 2025, Trondheim, Norway}
\acmDOI{10.1145/3696630.3728569} 
\acmISBN{979-8-4007-1276-0/2025/06} 

\usepackage{amsmath,amsfonts}
\usepackage{algorithmic}
\usepackage{graphicx}
\usepackage{textcomp}
\usepackage{multirow}
\usepackage{booktabs}
\usepackage{soul}
\usepackage{ulem}
\usepackage{diagbox} 
\usepackage[most]{tcolorbox}  
\usepackage{color}
\usepackage{xcolor}
\usepackage{tikz}
\usepackage{tikz-qtree}
\usepackage{pgfplots}
\usepackage{varwidth}
\usepackage[ruled,linesnumbered]{algorithm2e}
\usepackage{flushend}
\usetikzlibrary{patterns}
\usepackage{url}
\usepackage{subfigure}
\usepackage{threeparttable}
\usepackage{colortbl}
\usepackage{balance}
\usepackage{bbding}
\usetikzlibrary{plotmarks}
\usetikzlibrary{patterns}
\usepackage{pgfplots}
\usepackage{cleveref}
\usepackage{enumitem}
\usepackage{pifont} 

\begin{document}


\title{Towards Mitigating API Hallucination in Code Generated by LLMs with Hierarchical Dependency Aware}

\newcommand{\dataset}{APIHulBench}
\newcommand{\tool}{MARIN}
\newcommand{\modify}[1]{\textcolor{blue}{{#1}}}
\newcommand{\phaseone}{Hierarchical Dependency Mining}
\newcommand{\phasetwo}{Dependency Constrained Decoding}

\author{Yujia Chen}
\email{yujiachen@stu.hit.edu.cn}
\affiliation{%
  \institution{Harbin Institute of Technology}
  \city{Shenzhen}
  \country{China}
}

\author{Mingyu Chen}
\email{220110123@stu.hit.edu.cn}
\affiliation{%
  \institution{Harbin Institute of Technology}
  \city{Shenzhen}
  \country{China}
}

\author{Cuiyun Gao*}\thanks{* Corresponding author}
\email{gaocuiyun@hit.edu.cn}
\affiliation{%
  \institution{Harbin Institute of Technology}
  \city{Shenzhen}
  \country{China}
}

\author{Zhihan Jiang}
\email{jiangzhihan2@huawei.com}
\affiliation{%
  \institution{Huawei Cloud Computing Technologies Co., Ltd.}
  \city{Shenzhen}
  \country{China}
}

\author{Zhongqi Li}
\email{lizhongqi7@huawei.com}
\affiliation{%
  \institution{Huawei Cloud Computing Technologies Co., Ltd.}
  \city{Shenzhen}
  \country{China}
}

\author{Yuchi Ma}
\email{mayuchi1@huawei.com}
\affiliation{%
  \institution{Huawei Cloud Computing Technologies Co., Ltd.}
  \city{Shenzhen}
  \country{China}
}


\begin{abstract}
  Application Programming Interfaces (APIs) are crucial in modern software development. Large Language Models (LLMs) assist in automated code generation but often struggle with API hallucination, including invoking non-existent APIs and misusing existing ones in practical development scenarios. Existing studies resort to Retrieval-Augmented Generation (RAG) methods for mitigating the hallucination issue, but tend to fail since they generally ignore the structural dependencies in practical projects and do not indeed validate whether the generated APIs are available or not. To address these limitations, we propose {\tool}, a framework for mitigating API hallucination in code generated by LLMs with hierarchical dependency aware. {\tool} consists of two phases: \textit{{\phaseone}}, which analyzes local and global dependencies of the current function, aiming to supplement comprehensive project context in LLMs' input, and \textit{{\phasetwo}}, which utilizes mined dependencies to adaptively constrain the generation process, aiming to ensure the generated APIs align with the project's specifications. To facilitate the evaluation of the degree of API hallucination, we introduce a new benchmark {\dataset} and two new metrics including Micro Hallucination Number (MiHN) and Macro Hallucination Rate (MaHR). Experiments on six state-of-the-art LLMs demonstrate that {\tool} effectively reduces API hallucinations, achieving an average decrease of 67.52\% in MiHN and 73.56\% in MaHR compared to the RAG approach. Applied to Huawei's internal projects and two proprietary LLMs, {\tool} achieves average decreases of 57.33\% in MiHN and 59.41\% in MaHR.

\end{abstract}



\maketitle

\section{INTRODUCTION} \label{sec:intro}
    

\begin{figure*}
    \centering
    \includegraphics[scale=0.53]{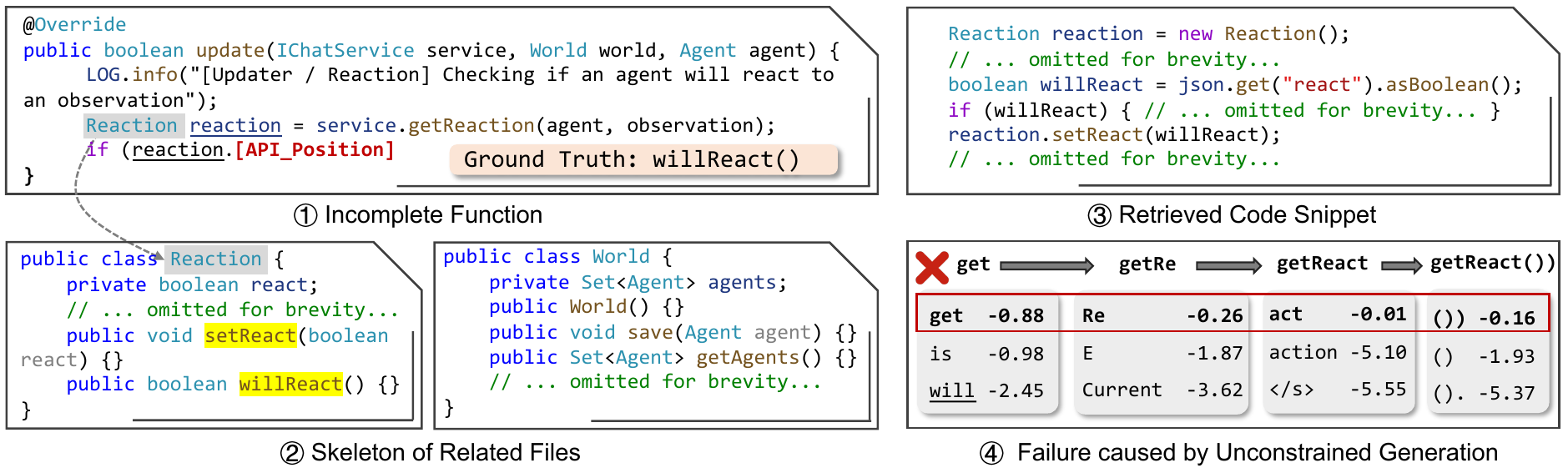}
    \caption{The motivation example: A Wrong API generated by CodeLLama-7B with RAG.}
    \label{fig:intro_example1}
\end{figure*}

In modern software development, developers frequently leverage application programming interfaces (APIs) to integrate complex functionalities into their projects efficiently~\cite{intro_api,intro_api_2,intro_api_3, DBLP:journals/tse/ChenGRP0L23,DBLP:conf/wcre/ChenGZ0WX24}. The advent of large language models (LLMs) has revolutionized this process, with AI-powered coding assistants greatly enhancing developers' productivity by generating code snippets automatically~\cite{intro_au_1,intro_au_2,intro_productivity_1,intro_productivity_2,intro_productivity_3}. However, despite their impressive capabilities, LLMs still face notable challenges in industrial development scenarios. A recent study evaluating six mainstream LLMs on practical code generation shows that 31.67\% of erroneous code directly results from incorrect API usage, a phenomenon named ``API hallucinations''~\cite{zhang2024llm}, including invoking non-existent APIs, misusing existing APIs, and 
incorrectly using arguments~\cite{jain2024mitigating}. Such mistakes increase debugging complexity since they can propagate errors in subsequent programming.
Moreover, a recent study reports that 29\% of security vulnerabilities in the Linux kernel in 2021 were traced back to improper API usage~\cite{intro-misuseapi}. Given the critical role of APIs in developers' daily workflows and the severe consequences of their misuse, exploring effective mitigation methods has become a critical research focus.
Existing studies primarily rely on Retrieval-Augmented Generation (RAG) techniques~\cite{jain2024mitigating,eghbali2024hallucinator,zhang2024llm}, which first construct the retrieval corpora including API documentation and code snippets from the project, and then integrate the retrieved relevant information into the input prompt.
While these studies have demonstrated success in mitigating API hallucination, they still face the following limitations. 

\textit{1) Ignoring structural dependencies in projects}. When developers call APIs, they typically rely on the structural dependencies of the current code snippet, such as relationships with other functions and modules, to ensure proper API usage within the project context. However, existing approaches fail to consider these critical dependencies, and instead, they offer isolated code snippets as references for LLMs. As illustrated in Figure~\ref{fig:intro_example1}\textcircled{1}, the incomplete function attempts to determine whether an agent will react, and the correct API to be used is {\tt willReact}. 
However, the retrieved code snippet (Figure~\ref{fig:intro_example1}\textcircled{3}) fails to provide sufficient context from related files (Figure~\ref{fig:intro_example1}\textcircled{2}) where the {\tt willReact} method is defined.
Besides, maintaining a retrieval corpus with updated indexing introduces additional resource costs, limiting the scalability of such approaches. 

\textit{2) Unconstrained API generation.} Existing approaches 
utilize auto-regressive decoding, where tokens are generated solely based on model probabilities without explicit constraints on the output space {of API tokens. This unconstrained process often leads to the generation of invalid API tokens. As demonstrated in Figure~\ref{fig:intro_example1}\textcircled{4}, the model generates candidate API tokens {\tt get}, {\tt Re}, {\tt act}, 
{\tt ())}, which does
not match the intended API {\tt willReact} defined in the {\tt Reaction} class. Such errors occur because the model lacks constraints to align generated tokens
with the actual project specifications, resulting in incorrect and infeasible API calls. This highlights the need to enforce output validity during the decoding process.

\noindent \textbf{Our work.} To address these limitations, in this paper, we propose {\tool}, a framework designed to \textbf{M}itigate \textbf{A}PI hallucination in code gene\textbf{R}ated by LLMs with h\textbf{I}erarchical depe\textbf{N}dency aware. Specifically, {\tool} comprises two phases: \textit{1) {\phaseone}.} This phase uses static analysis to analyze local and global dependencies of the current function, such as method call relationships and file dependencies, forming a structural
representation of the project to enrich LLM's input prompt. \textit{2) {\phasetwo.}} This phase adaptively constrains the token generation process using valid API patterns derived from the mined dependencies, ensuring the alignment of generated APIs
with project specifications. 
Through the two phases, {\tool} employs project structural dependencies both in supplying the input and guiding the generation, eliminating the need for retrieval corpora construction, thus enabling effective API hallucination mitigation in the code generation process of LLMs.

Existing API hallucination evaluation benchmarks suffer from potential data leakage from older repositories~\cite{eghbali2024hallucinator} or reliance on model-synthesized data~\cite{jain2024mitigating, tian2024codehalu}. Therefore, we introduce {\dataset}, a novel benchmark, comprising 416 high-quality samples from 98 recent Java repositories on GitHub. It encompasses two different coding processes: {\dataset}-F (221 samples) where developers initiate new functions, and {\dataset}-M (195 samples) where developers have written most of the code in functions. For evaluating the degree of API hallucination, we also propose two new metrics: Micro Hallucination Number (MiHN), which quantifies
the number of hallucinated elements within one generated API, and Macro Hallucination Rate (MaHR), which measures
the proportion of the generated APIs containing hallucinations. We apply {\tool} to six state-of-the-art LLMs with varying architectures and sizes (CodeLlama-7B/13B/34B~\cite{codellama} and DeepSeekCoder-1.3B/6.7B/33B~\cite{deepseek}) using the {\dataset} benchmark.
Experimental results demonstrate that {\tool} can significantly mitigate API hallucination of LLM-generated code, with an average decrease of 67.52\% in MiHN and 73.56\% in MaHR as well as an average increase of 107.3\% in exact match (EM) and 44.79\% in edit similarity (ES) compared to RAG approach. To further validate {\tool}'s effectiveness in industry settings, we construct an additional benchmark comprising 109 samples from Huawei's internal projects and apply {\tool} on Huawei's proprietary LLMs (PanguCoder-11B/34B~\cite{pangucoder}). 
The results indicate that {\tool} effectively and efficiently mitigates API hallucination, achieving an average decrease of 59.41\% in MaHR and an average increase of 72.39\% in EM, with only an average 0.031s overhead increase compared to the base model. We release {\dataset} at \url{https://github.com/yujiachen99/APIMitigation}. 

\definecolor{p1}{RGB}{174,223,172} 
\definecolor{p2}{RGB}{224,175,107}  
\definecolor{p3}{RGB}{138,170,214}  
\definecolor{p4}{RGB}{222,117,123} 
\definecolor{p5}{RGB}{216,174,174} 
\definecolor{p6}{RGB}{163,137,214} 
\definecolor{p7}{RGB}{248,199,1} 
\definecolor{p8}{RGB}{205,205,205} 
\definecolor{p9}{RGB}{255,0,127} 

\pgfplotstableread[row sep=\\,col sep=&]{
	datasets & evol & oss \\
	  2 & 0.1454 & 0.8272\\
        4 & 0.1090 & 0.8454\\
        6 & 0.1272 & 0.8727\\
        8 & 0.1181 & 0.8636\\
}\dskfive

\pgfplotstableread[row sep=\\,col sep=&]{
	datasets & base & RAG & P-C & API-G\\
	  2 & 0.8272 & 0.7272 & 0.6521 & 0.5636\\
        4 & 0.8454 & 0.7181 & 0.6884 & 0.5289\\
        6 & 0.8727 & 0.7181 & 0.6959 & 0.5818\\
        8 & 0.8636 & 0.7727 & 0.6884 & 0.5727\\
}\dskfive

\begin{figure*}[t]
\small
\setlength{\abovecaptionskip}{-2pt} 
\centering
\begin{minipage}[t]{0.475\textwidth} 
    \centering
    \subfigure[An illustration example of two types of APIs]{
        \includegraphics[width=\textwidth]{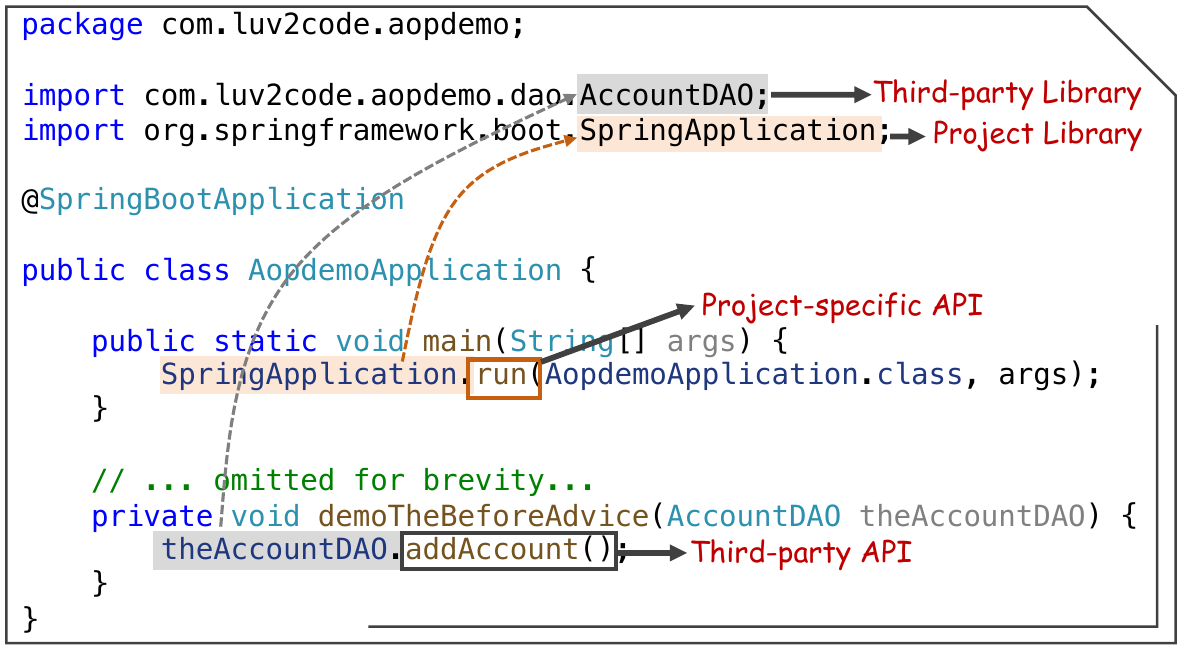}
        \label{fig:sub_a}
    }
\end{minipage}
\begin{minipage}[t]{0.48\textwidth} 
    \setlength{\abovecaptionskip}{-0.01cm}
    \setlength{\abovecaptionskip}{-25pt} 
    \centering
    \subfigure[Hallucination across different API types]{
        \begin{tikzpicture}[scale=0.75]
    \begin{axis}[
        grid=major,
        ybar=0.2pt,
        bar width=0.45cm,
        width=1.2\textwidth,
        height=0.35\textwidth,
        xtick=data,
        xticklabels={CL-7B, CL-13B, DSC-1.3B, DSC-6.7B},
        ymin=0, ymax=1.2,
        legend style={
            legend columns=-1, 
            draw=none, 
            font=\large, 
            at={(0.5,1.1)}, 
            anchor=south,
            inner sep=0pt
        },
        legend image code/.code={
            \draw [#1] (0cm,-0.08cm) rectangle (0.4cm,0.1cm); 
        },
        legend image post style={scale=0.9, thick},
        tick align=outside,
        xticklabel style={font=\normalsize},
        yticklabel style={font=\large},
        every axis plot/.append style={line width=1pt},
        every axis/.append style={line width=1pt},
        ylabel={\textbf{\large MaHR}}, 
        ticklabel style={font=\normalsize},
        nodes near coords,
        nodes near coords style={font=\large,xshift=-4pt}
    ]
    \addplot[fill=p7] table[x=datasets, y=evol]{\dskfive};
    \addplot[fill=p6] table[x=datasets, y=oss]{\dskfive};
    \legend{{\normalem Third-party APIs}, {\normalem Project-specific APIs}};
    \end{axis}
        \end{tikzpicture}
        \label{fig:sub_b}       
        \setlength{\belowcaptionskip}{-25pt} 
    }
    \vspace{-2mm} 
    \subfigure[Hallucination in different input contexts]{
    \setlength{\abovecaptionskip}{-25pt} 
        \begin{tikzpicture}[scale=0.75]
            \begin{axis}[
                grid=major,
                ybar=0.2pt,
                bar width=0.45cm,
                enlarge x limits=0.17,
                width=1.2\textwidth,
                height=0.35\textwidth,
                xtick=data,
                xticklabels={CL-7B, CL-13B, DSC-1.3B, DSC-6.7B},
                ymin=0.45, ymax=1.19,
                legend style={at={(0.95,1.25)}, anchor=east,legend columns=-1,draw=none,font=\large},
                legend image code/.code={
            \draw [#1] (0cm,-0.08cm) rectangle (0.4cm,0.1cm); },
                legend image post style={scale=0.9, thick},
                legend cell align={left}, 
                tick align=outside,
                xticklabel style={font=\normalsize},
                yticklabel style={font=\large},
                every axis plot/.append style={line width=1pt},
                every axis/.append style={line width=1pt},
                ylabel={\textbf{\large MaHR}}, 
                nodes near coords,
                nodes near coords style={font=\large,xshift=3pt}
            ]
            \addplot[fill=p1, nodes near coords, every node near coord/.append style={yshift=5pt}] table[x=datasets, y=base]{\dskfive};
            \addplot[fill=p2, nodes near coords, every node near coord/.append style={yshift=4pt}] table[x=datasets, y=RAG]{\dskfive};
            \addplot[fill=p3, nodes near coords, every node near coord/.append style={yshift=-1pt,xshift=-1pt}] table[x=datasets, y=P-C]{\dskfive};
            \addplot[fill=p4] table[x=datasets, y=API-G]{\dskfive};
            \legend{{\normalem Base}, {\normalem RAG}, {\normalem Project files}, {\normalem API reference}};
            \end{axis}
        \end{tikzpicture}
        \label{fig:sub_c}
        \setlength{\belowcaptionskip}{-25pt} 
    }
\end{minipage}
\caption{Analysis of API hallucinations across different dimensions.}
\label{fig:RQ4_CR}
\end{figure*}
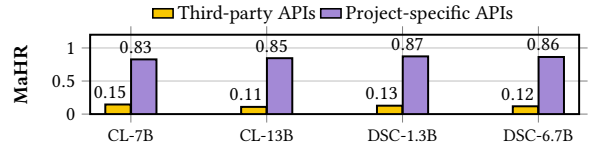
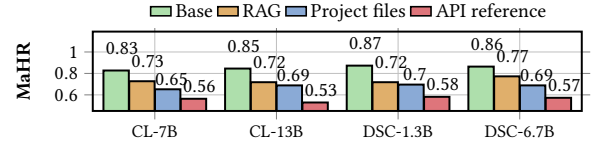

\noindent \textbf{Contributions.} The main contributions of our paper are summarized below:

\begin{itemize}[leftmargin=*]
     \item To the best of our knowledge, we are the first to systematically investigate API hallucination in the code generation process under practical scenarios and propose one benchmark and two metrics.

    \item We propose {\tool}, a novel framework that supplements hierarchical dependencies externally and constrains the decoding process internally, 
    mitigating API hallucination in the code generation process of LLMs.
    
    \item We evaluate {\tool} on both open-source and industrial scenarios. Experimental results show that our framework outperforms baselines, effectively and efficiently mitigating API hallucination and improving code generation quality.
    
\end{itemize}

\section{PRELIMINARY STUDY} \label{sec:motivation}
To better understand API hallucination in the industrial scenario, we conduct a preliminary study using a Huawei internal project. The study explores the prevalence of hallucinations across different API types and input contexts. We collect a dataset of 200 code snippets, including 100 third-party library API calls and 100 project-specific API calls, covering common API usage patterns in industrial software~\cite{api-type}, as shown in Figure~\ref{fig:sub_a}. The evaluation involves four widely-used LLMs, including CodeLlama-7B/13B~\cite{codellama} and DeepSeek-Coder-1.3B/6.7B~\cite{deepseek}, and a metric for measuring the portion of hallucinated APIs, named MaHR.

\subsection{Hallucination in Different API Types}

We input the collected code snippets into the four LLMs and analyze the proportion of hallucinated APIs. As shown in Figure~\ref{fig:sub_b}, the models generate few hallucinations for third-party library APIs, with an average rate of only 12.75\%. However, for project-specific APIs, the hallucination rate is much higher, averaging 85.25\%. This indicates that while models handle third-party library APIs well due to their frequent presence in training data, they struggle with project-specific APIs due to a lack of contextual awareness. 
Notably, larger models show only marginal improvements in reducing hallucinations compared to smaller models, indicating that increasing model size alone is insufficient to address this issue. 
These findings highlight the challenge of API hallucination in industrial scenarios, especially for project-specific APIs.

\subsection{Impact of Different Input Contexts}

To further explore ways to mitigate hallucination in project-specific APIs, we design four input context scenarios: (1) \textit{Base}, only using the collected code snippets; (2) \textit{RAG}~\cite{zhang2024llm}, using BM-25 to retrieve similar code snippets from the project; (3) \textit{Project files}, including all project files randomly; (4) \textit{API reference}, explicitly providing validated APIs as guidance. The results are shown in Figure~\ref{fig:sub_c}, and we make the following observations:

\begin{itemize}

    \item \textbf{Observation 1:} Including more project information (Scenario 3) improves performance and even outperforms the RAG-based approach (Scenario 2). This highlights the importance of a broader project context for API hallucination mitigation. However, due to context window limits, it is not feasible to include all project files.

    \item \textbf{Observation 2:}Including API references in the input context (Scenario 4) helps reduce some hallucinations, but the rates remain still very high, averaging 44\% across four models. This indicates that API references alone are insufficient to fully guide the model toward accurate API generation.
    
\end{itemize}

In summary, our study highlights the challenges of mitigating API hallucination, particularly for project-specific APIs. While a broader project context improves performance, it is limited by input constraints. Additionally, providing API information alone fails to reduce the hallucination issue.
These results emphasize the need for more effective approaches
to address API hallucination in industrial development.

\section{APPROACH} \label{sec:approach}
    \begin{figure*}[h]
    \centering
    \includegraphics[scale=0.52]{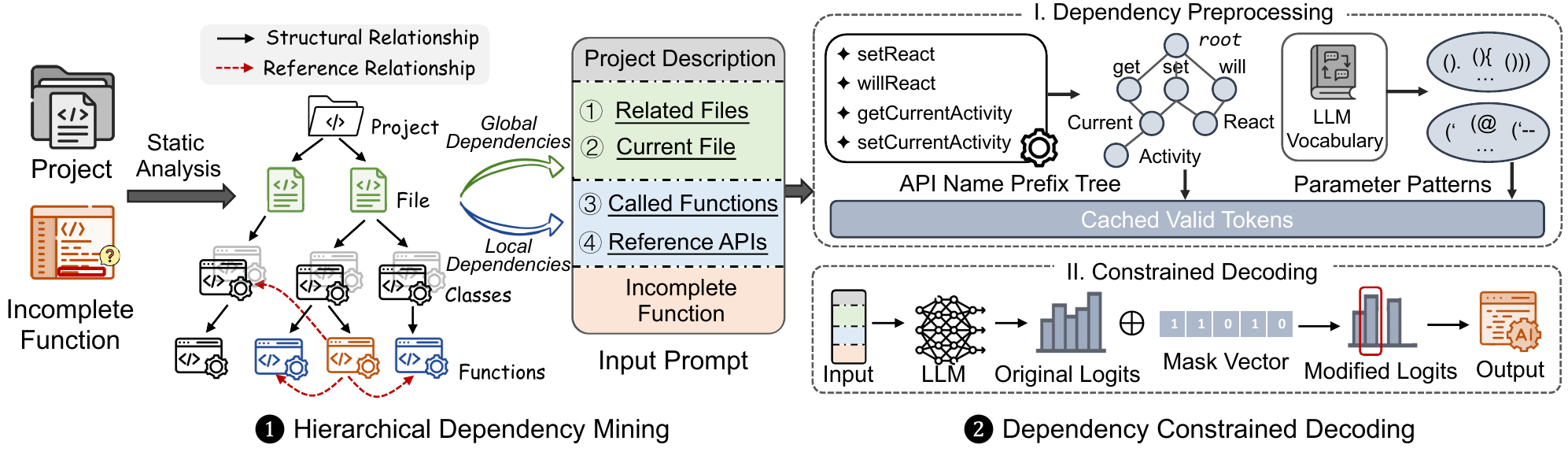}
    \caption{The overview of {\tool}.}
    \label{fig:framework}
\end{figure*}

In this section, we propose {\tool}, a framework to mitigate API hallucination in code generated by LLMs with hierarchical dependency aware. We first present the overview of {\tool} and then describe its details in the following subsections.

\subsection{Overview}

Given a project and an incomplete function, {\tool} generates a complete API call by a two-phase approach: (1) In the \textit{hierarchical dependency mining} phase, it extracts both local dependencies (i.e., called functions and reference APIs) and global dependencies (i.e., related files and the current file), enriching the input prompt to provide better context for LLM. (2) In the \textit{dependency constrained decoding} phase, the extracted dependencies are used to guide the generated API calls to align with the project's specifications by constructing valid API name prefix trees and identifying parameter patterns.


\subsection{{\phaseone}}

As shown in Figure~\ref{fig:framework} {\large\ding{182}}, {\tool} analyzes dependencies at two levels: local (function-level) and global (file-level), constructing a hierarchical structure of the project. This structure decomposes the project top-down, from files down to detailed functions. Beyond explicit structural relationships, it also incorporates implicit reference dependencies (See Section~\ref{sec:global}). Therefore, it provides a rich context for code generation that spans both immediate function details and broader project architecture.

\subsubsection{Local Dependency Analysis}

It focuses on the immediate context of the incomplete function. Through static analysis, we first identify valid APIs available at the generation position, providing clear references for the LLM's output. We also analyze method dependencies, including functions called within the same file or across other files in the project. This process captures both function calls and data flow, offering detailed relationships surrounding the incomplete function.

\subsubsection{Global Dependency Analysis} \label{sec:global}

While local dependencies provide immediate context, understanding broader project relationships is crucial for generating code within a large project, which is the global dependency. For the unfinished function, we first leverage static analysis to extract the import statements in its current file and then identify all imported files. These related files provide essential context about the available methods for the current function. To avoid overwhelming LLMs and exceeding input length limits, we extract a concise ``skeleton'' of these files. Specifically, for each file, we preserve its class definitions, field declarations, and function signatures while excluding detailed implementations. This approach provides a compact yet informative representation of the global dependency on projects.




\begin{figure}
    \centering
    \setlength{\belowcaptionskip}{-3pt} 
    \includegraphics[scale=0.47]{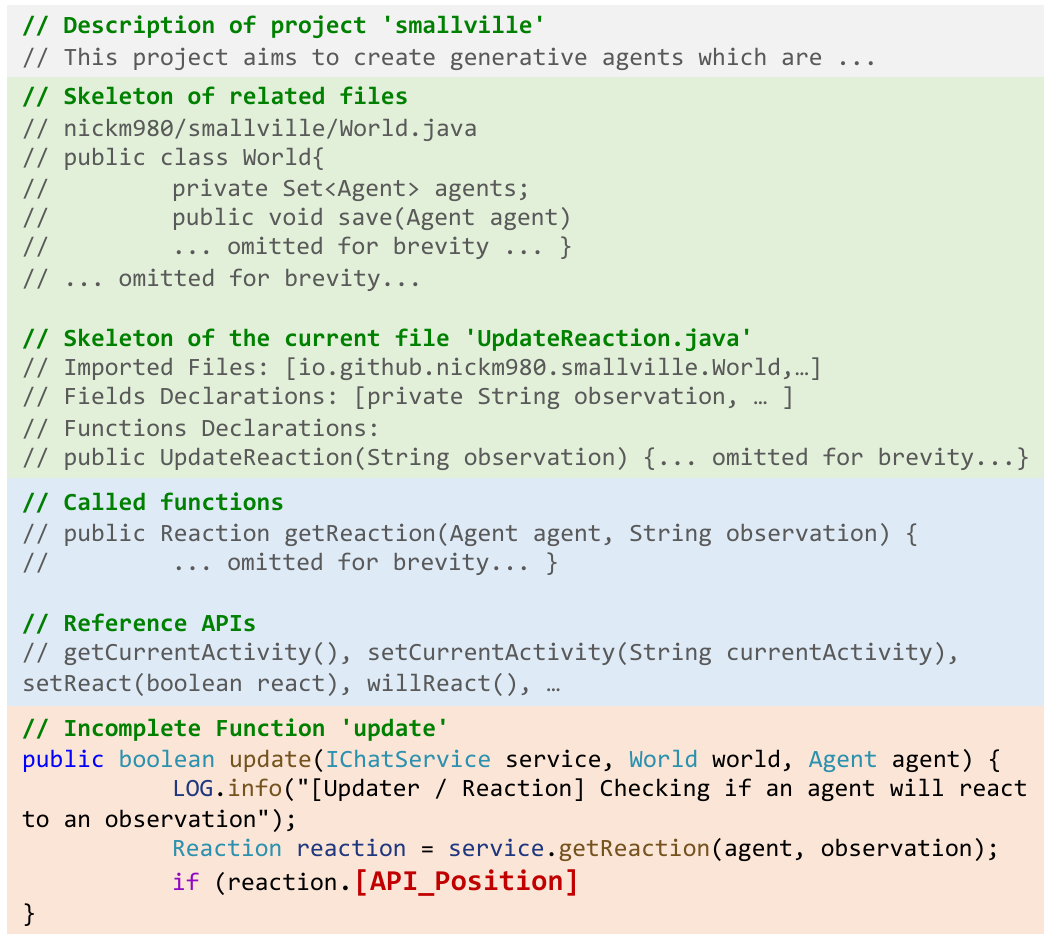}
    \caption{An example illustrating the {\tool}' prompt template based on the incomplete function in Section~\ref{sec:intro}.}
    \label{fig:prompt}
\end{figure}
\subsubsection{Input Prompt Construction}

Based on the dependency analysis, we construct a structured input prompt, as illustrated in Figure~\ref{fig:prompt}. This prompt integrates the following key components:

\begin{itemize}[leftmargin=*]
    \item \textbf{Project Description.} A brief overview of the project's purpose, providing essential background for the LLM.

    \item \textbf{Global Dependency.} Simplified skeletons of related and current
    files, including class definitions, member fields, and function signatures, to represent the broader project structure.

    \item \textbf{Local Dependency.} Called functions and reference APIs that are directly relevant to the incomplete function, serving as 
    its immediate context.
    
    \item \textbf{Incomplete Function.} The partially implemented target function, with the {\tt[API\_Position]} marking the location where the API call should be generated.
    
\end{itemize}

\subsection{{\phasetwo}}

This phase first analyzes the mined dependencies to preprocess valid API-related information and then integrates these constraints into the decoding process, as illustrated in Figure~\ref{fig:framework}{\large\ding{183}}. 

\subsubsection{Dependency Preprocessing} In this step, dependencies are used to build an API name prefix tree and identify parameter patterns, forming a constrained set of valid options. These valid tokens are cached to ensure efficient access during decoding, as shown in Figure~\ref{fig:framework}{\large\ding{183}-I}. 


\textbf{API name prefix tree.} Given a reference API set $\mathcal{A}=\{a_1, a_2, \cdots a_n\}$, 
we tokenize the name of each API $a_i$ to $\mathcal{T}(a_i)$ based on the LLM's vocabulary. It is
important to note that tokenization in LLMs is context-dependent. For example, in CodeLlama-7B, the tokenization sequence for {\tt willReact} is [674, 1123, 627], but in the context of {\tt reaction.willReact}, it becomes [14043, 1123, 627]. To ensure consistent tokenization, we tokenize both the full API calls and their prefixes (i.e., the tokens before the last ``.''). We then remove the shared prefixes, yielding a deterministic representation $\mathcal{T}(a_i)$ for each API. In real-world projects, reference APIs often share common prefixes. For efficient constraint validation during generation, we construct a prefix tree from these APIs' token sequences, as described in Algorithm~\ref{alg:trie}. The algorithm initializes the tree with an empty root node (line 1) and iterates through each reference API (line 2). For each token in the tokenized API, it builds the tree recursively (lines 3-8): if a token is not a child of the current node, a new node is created (lines 4-5), then the algorithm moves to this child node (line 7). Upon completion, each root-to-leaf path in the tree represents a complete API token sequence, enabling efficient validation of generated API calls.


\begin{algorithm}[t]
  \caption{Prefix Tree Construction for API Tokens}
  \label{alg:trie}
  \small
  \SetKwInOut{Input}{input}\SetKwInOut{Output}{output}
  \Input{Reference APIs $\mathcal{A}$, Tokenizer $\mathcal{T}$}
  \Output{Prefix tree $G$}
    Initialize the prefix tree $G$ with a root node $\mathcal{R}$, where $\mathcal{R}$.child $\gets \emptyset$\;
  \ForEach{ $a_i \in \mathcal{A}$}{
  \tcp{\textcolor{teal}{recursively build the prefix tree;}}
      \ForEach{$t_i \in \mathcal{T}(a_i)$}{
          \If{$t_i \notin \mathcal{R}$.child}{
              \tcp{\textcolor{teal}{add $t_i$ into the children of $\mathcal{R}$;}}
              $\mathcal{R}.$child[$t_i$] $\gets$ create a new node with $t_i$\;
          }
        $\mathcal{R}\gets \mathcal{R}$.child[$t_i$]\;
        
      }
  }
  \Return{$\mathcal{R}$}\;

\end{algorithm}




\textbf{Parameter Patterns.} We use an indicator function $I$ to differentiate between APIs with and without parameters. To handle parameter-related tokens in the LLM vocabulary, we classify them into two categories:

\begin{itemize}[leftmargin=*]
    \item $B_{noparam}$: Tokens representing no-parameter patterns (e.g., ``())'', ``()))\{'') that appear when an API is called without arguments.

     \item $B_{param}$: Tokens representing parameter patterns (e.g., ``([\text{--}'', ``(@'') that indicate the start of parameter lists.
\end{itemize}

\subsubsection{Constrained Decoding} As shown in Figure~\ref{fig:framework}{\large\ding{183}-II}, at each step $t$, the LLM outputs logits $l \in \mathbb{R}^{|V|} $ representing token probabilities over its vocabulary $V$. To ensure the generated APIs conform to mined dependencies, we employ an adaptive 
masking mechanism that combines both API name and parameter pattern constraints. 

Given the current sequence $\{t_1, t_2, ..., t_k\}$, if $t_{k+1}$ is an API name token, we locate the current node in the trie structure and collect valid next tokens from its children as $V_{valid} = \{ v| v \in n.children\}$. if $t_{k+1}$ is a parameter token,  we consult the parameter indicator $I$ to determine valid tokens - allowing only no-parameter tokens when $I=0$ or parameter tokens when $I=1$.
%
These constraints are integrated into a unified binary mask $M \in \{  0,1\}^{|V|}$:

\begin{equation}
M_i = \begin{cases}
1 & \text{if } i \in V_{valid} \text{ (API name constraint)} \\
1 & \text{if } i \in B_{noparam} \text{ and } I = 0 \text{ (no-param constraint)} \\
1 & \text{if } i \in B_{param} \text{ and } I = 1 \text{ (param constraint)} \\
0 & \text{otherwise}
\end{cases}
\end{equation}

The constrained logits are then computed through element-wise multiplication $l_{masked} = l \odot M$, and the next token is selected via greedy search: 
\begin{equation}
 t_{(k+1)}=argmax_t(softmax(l_{masked})).
\end{equation}
This process iterates until a complete API call is generated or a maximum length is reached.

\section{EXPERIMENTAL SETUP} \label{sec:setup}
    
\subsection{Research Questions} \label{sec:RQ}

In this paper, we mainly investigate the following research questions through experiments.

\begin{itemize}[leftmargin=*]

\item {\textbf{RQ1:}} How effective is {\tool} in mitigating API hallucination on {\dataset}?
\item {\textbf{RQ2:}} What are the impacts of two components in {\tool} (i.e., \textit{hierarchical dependency mining} and \textit{dependency constrained decoding})?
\item {\textbf{RQ3:}} How efficient is {\tool} compared to baseline approaches?
\item {\textbf{RQ4:}} How does {\tool} perform in mitigating API hallucination in an industrial scenario?
\end{itemize}

\subsection{Benchmarks} \label{sec:benchmark}


Existing API hallucination evaluation benchmarks face challenges such as potential data leakage from older projects~\cite{eghbali2024hallucinator} and reliance on model-synthesized data~\cite{tian2024codehalu,jain2024mitigating}. To address these issues, we construct a new benchmark, {\dataset}, which focuses on project-specific API calls in real development scenarios, providing a more practical evaluation of LLMs.

\subsubsection{Project Collection}
We collect Java projects from GitHub created between mid-2023 and mid-2024, ensuring they postdate the training cut-off dates of many existing code LLMs~\cite{benchmark-time}. To ensure quality, following prior work~\cite{bench_year}, we excluded projects with fewer than 10 files or fewer than 200 stars, resulting in a final dataset of 98 high-quality projects.

\subsubsection{Dataset Generation}
For each project, we use tree-sitter~\cite{tree-sitter} to parse Java files and extract functions with more than five lines as candidate samples. We identify project-specific API calls within these functions and split each function into two parts: the prompt (code before the API) and an inference part (containing the ground truth API). The line position of each API call is recorded. Following prior studies~\cite{benchmark-time,bench_year}, we eliminate duplicate prompts, resulting in a benchmark with 416 unique samples. 
To reflect different development stages, we divide the benchmark into two variants based on API call positions: 1) {\dataset}-F, including 221 samples where API calls appear in the first 50\% of the function lines, reflecting scenarios where developers are initiating their implementation and seeking early assistance. 2) {\dataset}-M, including 195 samples where API calls appear beyond the 50\% mark, simulating scenarios where developers have written most of the code and need assistance during later stages of development.
This division enables a more nuanced evaluation of LLM performance across different stages of coding, providing insights into their practical performance in real-world development.

\subsection{Studied LLMs} 

Following prior studies~\cite{wavecoder,Magicoder,zhang2024llm}, we adopt two widely-used LLMs for code generation with different model sizes:

\begin{itemize}[leftmargin=*]
    \item \textbf{CodeLlama~\cite{codellama}} released by Meta AI, is further trained on Llama 2~\cite{llama2} with 0.5 trillion code-specific tokens. Our experiments use its 7B, 13B, and 34B versions.
    \item \textbf{DeepSeekCoder~\cite{deepseek}} released by DeepSeek AI, is a series of code-specialist LLMs trained on 2 Trillion tokens across over 80 programming languages. It achieves state-of-the-art performance among open-source code LLMs. Our experiments use its 1.3B, 6.7B, and 33B versions.
\end{itemize}

\subsection{Studied Baselines} \label{sec:baseline}

To evaluate the effectiveness of {\tool}, we compare it with the following baselines. Note that we do not include DAG~\cite{jain2024mitigating} in the comparison because this approach requires detailed documentation for each API, which is unavailable for many projects.

\begin{itemize}[leftmargin=*]

\item \textbf{Base Generation} is a default generation strategy using the current function as the input prompt.

\item \textbf{RAG~\cite{zhang2024llm}} retrieves code snippets from the project similar to the incomplete function and includes them in the input prompt.

\item \textbf{De-Hallucinator~\cite{eghbali2024hallucinator}} first generates code using the incomplete function. Then, it retrieves APIs related to this initial output, adds them to the input prompt, and iteratively refines the generation.

\end{itemize}

\subsection{Metrics}

To comprehensively evaluate the performance of LLMs, we focus on two aspects: generation accuracy and generation hallucination.

\subsubsection{Accuracy}

Following prior work~\cite{eghbali2024hallucinator}, we adopt three metrics to evaluate generation accuracy:

\begin{itemize}[leftmargin=*]

\item \textbf{Exact Match (EM)} measures the percentage of output that exactly matches the ground truth.

\item \textbf{Edit Similarity (ES)} calculates the Levenshtein distance between the output and the ground truth, normalized to a similarity score.

\item \textbf{Identifier Match (IM)} measures the percentage of output with correct API identifiers compared to the ground truth.

\end{itemize}

\subsubsection{Hallucination}

We focus on two key elements in APIs, including the 
names and
parameter patterns, and
introduce two new metrics for evaluating the degree of API hallucination.


\begin{itemize}[leftmargin=*]

\item \textbf{Micro Hallucination Number (MiHN)} counts the average number of hallucinatory elements within the generated APIs:

\begin{equation}
\text{MiHN} = \frac{\sum_{i=1}^{n} \text{Count}(hallucinatory \ elements \ in \ a_i)} {n},
\end{equation}
where $n$ is the total number of generated APIs, and $a_i$ is the $i$-th generated API.

\item \textbf{Macro Hallucination Rate (MaHR)} calculates the proportion of generated APIs that contain any hallucinatory elements:

\begin{equation}
\text{MaHR} = \frac{\text{Count}(APIs \ with \ hallucinations)} {n},
\end{equation}
where $n$ is the total number of generated APIs.

\end{itemize}

\subsection{Implementation Details}

Given the limited input size of 8,000 tokens, we restrict the project content to 7,000 tokens and the function content to 1,000 tokens (as the average length of incompetent function in our benchmark is 358). Content exceeding these limits is truncated from left to right, which ensures adequately and equally sized input windows for different baselines. The prompts of RAG~\cite{zhang2024llm} and De-Hallucinator~\cite{eghbali2024hallucinator} are constructed following their original papers. For a fair comparison, we run De-Hallucinator one iteration. We download all LLMs from HuggingFace Hub~\cite{HuggingFace}, use vllms~\cite{vllms} on Ascend as the inference and serving engine, and set the max\_new\_tokens parameter to 15. All experiments are conducted on a server with 192-kunpeng-920 CPU cores, 1.5T memory, hosted by EulerOS 2.1 Linux distribution, and equipped with 4 Ascend 910B-B3 64G NPUs. 


\section{RESULTS} \label{sec:results}

\begin{table*}[t]
\caption{Evaluation results of baseline approaches and {\tool} on two benchmarks, \textbf{\raisebox{0.5ex}{\textasteriskcentered}} denotes statistically significant improvement (t-test with p-value < 0.001) over the baseline approaches.}
\label{Table:res_main}
\begin{tabular}{c|c|ccccc|ccccc}
\toprule
\multirow{2}{*}{\textbf{Model}} & \multirow{2}{*}{\textbf{Approach}} & \multicolumn{5}{c|}{\textbf{{\dataset}-F}} & \multicolumn{5}{c}{\textbf{{\dataset}-M}} \\ \cline{3-12} 
 &  & EM(\%) & ES(\%) & IM(\%) & MaHR(\%) & MiHN & EM(\%) & ES(\%) & IM(\%) & MaHR(\%) & MiHN  \\ \midrule
\multirow{4}{*}{CodeLlama-7B} & Base & 21.71 & 46.32 & 28.95 & 73.30 & 0.79 & 27.69 & 55.69 & 35.89 & 65.12 & 0.74 \\
 & RAG & 27.60 & 50.71 & 34.84 & 67.87 & 0.71 & 36.41 & 61.10 & 47.69 & 53.84 & 0.61 \\
 & De-Hallucinator & 22.17 & 46.79 & 29.41  & 73.30 & 0.81 & 29.89  & 56.62 & 40.20 &60.30  &  0.71 \\
& {\cellcolor{gray!20}\textbf{\tool}} & {\cellcolor{gray!20}\textbf{76.01\raisebox{0.5ex}{\textasteriskcentered}}} & {\cellcolor{gray!20}\textbf{87.04\raisebox{0.5ex}{\textasteriskcentered}}} & {\cellcolor{gray!20}\textbf{87.33\raisebox{0.5ex}{\textasteriskcentered}}} & {\cellcolor{gray!20}\textbf{15.83\raisebox{0.5ex}{\textasteriskcentered}}}  & {\cellcolor{gray!20}\textbf{0.17\raisebox{0.5ex}{\textasteriskcentered}}}  & {\cellcolor{gray!20}\textbf{66.66\raisebox{0.5ex}{\textasteriskcentered}}} & {\cellcolor{gray!20}\textbf{83.66\raisebox{0.5ex}{\textasteriskcentered}}} & {\cellcolor{gray!20}\textbf{89.23\raisebox{0.5ex}{\textasteriskcentered}}} & {\cellcolor{gray!20}\textbf{12.30\raisebox{0.5ex}{\textasteriskcentered}}} & {\cellcolor{gray!20}\textbf{0.13\raisebox{0.5ex}{\textasteriskcentered}}} \\ \midrule
\multirow{4}{*}{CodeLlama-13B} & Base & 33.03 & 60.05 & 39.81 & 61.08 & 0.71 & 33.33 & 61.73 & 41.02 & 59.48 & 0.70 \\
 & RAG & 49.32 & 69.78 & 59.27 & 42.98 & 0.48 & 36.41  & 61.61 & 48.71 & 52.82 &  0.61 \\
 & De-Hallucinator & 41.62 & 66.35 & 49.32 & 52.03 & 0.60 & 32.98 & 62.24 & 41.75 & 58.76 & 0.69 \\
& {\cellcolor{gray!20}\textbf{\tool}} & {\cellcolor{gray!20}\textbf{66.96\raisebox{0.5ex}{\textasteriskcentered}}} & {\cellcolor{gray!20}\textbf{80.20\raisebox{0.5ex}{\textasteriskcentered}}} & {\cellcolor{gray!20}\textbf{74.20\raisebox{0.5ex}{\textasteriskcentered}}} & {\cellcolor{gray!20}\textbf{28.95\raisebox{0.5ex}{\textasteriskcentered}}} & {\cellcolor{gray!20}\textbf{0.33\raisebox{0.5ex}{\textasteriskcentered}}} & {\cellcolor{gray!20}\textbf{66.66\raisebox{0.5ex}{\textasteriskcentered}}} & {\cellcolor{gray!20}\textbf{84.09\raisebox{0.5ex}{\textasteriskcentered}}} & {\cellcolor{gray!20}\textbf{88.20\raisebox{0.5ex}{\textasteriskcentered}}} & {\cellcolor{gray!20}\textbf{12.30\raisebox{0.5ex}{\textasteriskcentered}}} & {\cellcolor{gray!20}\textbf{0.13\raisebox{0.5ex}{\textasteriskcentered}}} \\ \midrule
\multirow{4}{*}{CodeLlama-34B} & Base & 19.45 & 51.62 & 27.14 & 75.56 & 0.81 & 24.61 & 55.02 & 32.30 & 68.71 & 0.74 \\
 & RAG & 31.67 & 55.21 & 39.36 & 63.80 & 0.66 & 38.46  & 62.59 & 48.20 & 52.82  & 0.58 \\
 & De-Hallucinator & 18.09 & 49.25 &25.79  & 76.47 & 0.84 & 26.28 & 55.50 & 33.50 & 67.52 & 0.76 \\
& {\cellcolor{gray!20}\textbf{\tool}} & {\cellcolor{gray!20}\textbf{81.90\raisebox{0.5ex}{\textasteriskcentered}}} & {\cellcolor{gray!20}\textbf{89.75\raisebox{0.5ex}{\textasteriskcentered}}} & {\cellcolor{gray!20}\textbf{90.49\raisebox{0.5ex}{\textasteriskcentered}}} & {\cellcolor{gray!20}\textbf{12.66\raisebox{0.5ex}{\textasteriskcentered}}} & {\cellcolor{gray!20}\textbf{0.13\raisebox{0.5ex}{\textasteriskcentered}}} & {\cellcolor{gray!20}\textbf{70.76\raisebox{0.5ex}{\textasteriskcentered}}} & {\cellcolor{gray!20}\textbf{86.24\raisebox{0.5ex}{\textasteriskcentered}}} & {\cellcolor{gray!20}\textbf{90.76\raisebox{0.5ex}{\textasteriskcentered}}} & {\cellcolor{gray!20}\textbf{10.76\raisebox{0.5ex}{\textasteriskcentered}}} & {\cellcolor{gray!20}\textbf{0.12\raisebox{0.5ex}{\textasteriskcentered}}} \\ \midrule
\multirow{4}{*}{DeepSeekCoder-1.3B} & Base & 21.71 & 52.19 & 29.41 & 72.85 & 0.87 & 23.58 & 52.71 & 32.82 & 68.20 & 0.82 \\
 & RAG & 38.91 & 63.04 & 46.15 & 56.10 & 0.67 & 34.87 & 61.26 & 44.61 & 56.41 & 0.71 \\
 & De-Hallucinator & 40.27 & 64.81 & 46.60 & 54.29 & 0.63 & 30.92 & 57.06 & 40.20 & 60.30 & 0.71 \\
& {\cellcolor{gray!20}\textbf{\tool}} & {\cellcolor{gray!20}\textbf{75.11\raisebox{0.5ex}{\textasteriskcentered}}} & {\cellcolor{gray!20}\textbf{86.82\raisebox{0.5ex}{\textasteriskcentered}}} & {\cellcolor{gray!20}\textbf{84.61\raisebox{0.5ex}{\textasteriskcentered}}} & {\cellcolor{gray!20}\textbf{17.64\raisebox{0.5ex}{\textasteriskcentered}}} & {\cellcolor{gray!20}\textbf{0.19\raisebox{0.5ex}{\textasteriskcentered}}} & {\cellcolor{gray!20}\textbf{64.61\raisebox{0.5ex}{\textasteriskcentered}}} & {\cellcolor{gray!20}\textbf{82.34\raisebox{0.5ex}{\textasteriskcentered}}} & {\cellcolor{gray!20}\textbf{85.64\raisebox{0.5ex}{\textasteriskcentered}}} & {\cellcolor{gray!20}\textbf{15.38\raisebox{0.5ex}{\textasteriskcentered}}} & {\cellcolor{gray!20}\textbf{0.16\raisebox{0.5ex}{\textasteriskcentered}}} \\ \midrule
\multirow{4}{*}{DeepSeekCoder-6.7B} & Base & 20.81 & 49.00 & 25.33 & 75.56 & 0.82 & 23.07  & 51.78 & 31.28 & 68.71 &  0.76 \\
 & RAG & 37.10  & 59.57 & 42.98 & 57.46 & 0.64 & 33.33 & 58.90 & 44.10 & 56.92 & 0.65 \\
 & De-Hallucinator & 31.22 & 56.88 & 39.36 & 64.25 &  0.72 & 24.22 & 53.39 & 32.47 & 67.52 & 0.77 \\
& {\cellcolor{gray!20}\textbf{\tool}} & {\cellcolor{gray!20}\textbf{75.56\raisebox{0.5ex}{\textasteriskcentered}}} & {\cellcolor{gray!20}\textbf{87.65\raisebox{0.5ex}{\textasteriskcentered}}} & {\cellcolor{gray!20}\textbf{86.42\raisebox{0.5ex}{\textasteriskcentered}}} & {\cellcolor{gray!20}\textbf{16.74\raisebox{0.5ex}{\textasteriskcentered}}} & {\cellcolor{gray!20}\textbf{0.18\raisebox{0.5ex}{\textasteriskcentered}}} & {\cellcolor{gray!20}\textbf{68.20\raisebox{0.5ex}{\textasteriskcentered}}} & {\cellcolor{gray!20}\textbf{85.23\raisebox{0.5ex}{\textasteriskcentered}}} & {\cellcolor{gray!20}\textbf{89.74\raisebox{0.5ex}{\textasteriskcentered}}} & {\cellcolor{gray!20}\textbf{11.28\raisebox{0.5ex}{\textasteriskcentered}}} & {\cellcolor{gray!20}\textbf{0.11\raisebox{0.5ex}{\textasteriskcentered}}} \\ \midrule
\multirow{4}{*}{DeepSeekCoder-33B} & Base & 35.74 & 57.99 & 41.62 & 59.72 & 0.66 & 20.00 & 51.98 & 30.25 & 70.25 & 0.78 \\
 & RAG & 46.15 & 67.40 & 54.29 & 47.05 & 0.50 & 36.41 &  61.90 & 46.15 & 54.35 & 0.60 \\
 & De-Hallucinator & 42.98 & 64.45 & 48.86 & 52.03 & 0.57 & 27.83 & 57.26 & 36.59 & 63.40  & 0.71 \\
& {\cellcolor{gray!20}\textbf{\tool}} & {\cellcolor{gray!20}\textbf{81.90\raisebox{0.5ex}{\textasteriskcentered}}} & {\cellcolor{gray!20}\textbf{90.83\raisebox{0.5ex}{\textasteriskcentered}}} & {\cellcolor{gray!20}\textbf{91.40\raisebox{0.5ex}{\textasteriskcentered}}} & {\cellcolor{gray!20}\textbf{11.76\raisebox{0.5ex}{\textasteriskcentered}}} & {\cellcolor{gray!20}\textbf{0.13\raisebox{0.5ex}{\textasteriskcentered}}} & {\cellcolor{gray!20}\textbf{67.69\raisebox{0.5ex}{\textasteriskcentered}}} & {\cellcolor{gray!20}\textbf{85.39\raisebox{0.5ex}{\textasteriskcentered}}} & {\cellcolor{gray!20}\textbf{91.28\raisebox{0.5ex}{\textasteriskcentered}}} & {\cellcolor{gray!20}\textbf{9.74\raisebox{0.5ex}{\textasteriskcentered}}} & {\cellcolor{gray!20}\textbf{0.10\raisebox{0.5ex}{\textasteriskcentered}}} \\ \bottomrule
\end{tabular}
\end{table*}

\subsection{Effectiveness of {\tool} on {\dataset} (RQ1)}

\noindent \textbf{Experimental Design.} To answer this research question, we conduct experiments on the two benchmarks {\dataset}-F and {\dataset}-M using six different LLMs.

\noindent \textbf{Results.} Table~\ref{Table:res_main} shows that {\tool} consistently outperforms baseline approaches across all evaluated LLMs. 
Using the Wilcoxon signed-rank test~\cite{wilcoxon1992individual} (p-value < 0.001), we confirm the improvements are statistically significant.

\textbf{{\tool} can be tailored to various programming scenarios.} 
The available content of incomplete function affects the base models' performance. For example, the six models achieve an average ES of 54.81\% and MaHR of 66.74\% on {\dataset}-M, while on {\dataset}-F, these metrics are 52.86\% and 69.67\%, respectively. {\tool} demonstrates strong performance in mitigating API hallucination across both scenarios. On {\dataset}-F, {\tool} improves the performance of the CodeLlama family by up to 175.39\% in EM while reducing MaHR by up to 80.15\% compared to RAG. On {\dataset}-M, similar gains are observed, with increasing up to 83.98\% in EM and reducing up to 79.62\% in MaHR. 
These results show {\tool} is well-suited for early-stage development with sparse context to later-stage tasks with richer dependencies.

\textbf{{\tool} shows generalization capability across various LLMs.} For DeepSeekCoder-1.3B, {\tool} boosts IM by 81.56\%, while reducing MiHN by 69.84\% compared to De-Hallucinator on {\dataset}-F. For larger models, DeepSeekCoder-33B, the impact remains substantial, improving IM by 87.06\% and reducing MiHN by 77.19\%. Similar trends are observed in the CodeLlama family. This generalization capability ensures {\tool} can be widely applied across different model architectures and sizes in diverse development scenarios.

\begin{tcolorbox}[breakable,width=\linewidth-2pt,boxrule=0pt,top=2pt, bottom=2pt, left=4pt,right=4pt, colback=gray!15,colframe=gray!15]
\textbf{Answer to RQ1:} {\tool} significantly mitigates API hallucination across diverse development scenarios and model sizes, achieving an average increase of 107.32\% in EM and an average decrease of 67.31\% in MaHR compared to the RAG approach.
\end{tcolorbox}

\input{datas/RQ2_figure}
\begin{table*}[t]
\caption{Computational time comparison across baseline approaches and {\tool}. The abbreviation ``De-Hal.'' denotes ``De-Hallucinator'', the ``CL'' denotes ``CodeLlama'' and the ``DSC'' denotes ``DeepSeekCoder''.}
\label{table:res_efficiency}
\begin{tabular}{c|ccc|ccc|ccc|ccc}
\toprule
\multirow{2}{*}{\textbf{Approach}} & \multicolumn{6}{c|}{\textbf{{\dataset}-F}}                       & \multicolumn{6}{c}{\textbf{{\dataset}-M}}                        \\ \cline{2-13} 
                          & CL-7B & CL-13B & CL-34B & DSC-1.3B & DSC-6.7B & DSC-33B & CL-7B & CL-13B & CL-34B & DSC-1.3B & DSC-6.7B & DSC-33B \\ \midrule
Base                      & 0.203      &  0.228   & 0.379       &   0.214       & 0.217         & 0.309          & 0.204      &  0.216      &  0.380      &   0.209       & 0.211         &   0.319      \\ \midrule
RAG               & 0.526  &  0.684  & 0.874    & 0.659 &  0.664 & 0.888  &  0.910  & 0.935  &   1.054   &      0.916    &   0.919       & 1.045         \\ \midrule
De-Hal.           &  0.571     &  0.729     & 0.906   &  0.551       &  0.773      & 0.865           &  0.560   &0.744      & 0.950        & 0.552   &  0.768        & 0.858     \\ \midrule
\rowcolor{gray!20} 
{\textbf{\tool}}         & \textbf{0.214}     &   \textbf{0.276}     &  \textbf{0.418}      &  \textbf{0.226}        &   \textbf{0.233}       &  \textbf{0.329}        &  \textbf{0.216}     &  \textbf{0.232}      & \textbf{0.453}       &  \textbf{0.210}        &  \textbf{0.215}        & \textbf{0.333}          \\ \bottomrule
\end{tabular}
\end{table*}


\subsection{Impacts of Different Components in {\tool} (RQ2)}

\noindent \textbf{Experimental Design.} For this RQ, we perform ablation studies by considering different levels of dependencies with and without constrained decoding (CD), resulting in eight variants.

\begin{itemize}[leftmargin=*] 
\item \textbf{ALL w/wo CD:} Provides complete hierarchical dependencies, including both local and global dependencies. 
\item \textbf{-LD w/wo CD:} Removes local dependencies, providing only related files and the current file. 
\item \textbf{-GD w/wo CD:} Removes global dependencies, providing only called functions and reference APIs. 
\item \textbf{-LG w/wo CD:} Removes both local and global dependencies, providing only the incomplete function. \end{itemize}

\noindent \textbf{Results.} Figures~\ref{fig:res_aba_F} and \ref{fig:res_aba_m} show the performance of eight variants on {\dataset}-F and {\dataset}-M, respectively. The results confirm that both hierarchical dependencies and constrained decoding independently improve the performance of {\tool}, with their combination achieving the best results.

\textbf{Global dependency is more beneficial for hallucination mitigation.} Removing global dependencies (-GD) causes a larger performance drop. For example, on {\dataset}-F, removing global dependencies causes a 33.39\% drop in average EM and a 186\% rise in average MaHR across six LLMs. In contrast, removing local dependencies results in a 15.79\% decrease in average EM and an 81.66\% increase in average MaHR.
This demonstrates that global dependencies, such as file structures and method signatures, are crucial for providing the context required for accurate API generation.

\textbf{Constrained decoding consistently improves hallucination mitigation across different input contexts.} For complete dependencies (ALL), with constrained decoding, on {\dataset}-F, the average EM increases by 9.12 points and MaHR decreases by 11.53 points across six LLMs. On {\dataset}-M, the EM increases by 11.28 points and MaHR decreases by 12.22 points. Even in limited dependency settings (-LD, -GD, -LG), constrained decoding also achieves the improvements. For example, in ``-LG'' variant, constrained decoding improves the EM by an average of 8.52 points and reduces MaHR by an average of 16.13 points on {\dataset}-F across six LLMs.
These results highlight the effectiveness of constrained decoding could employ the available dependencies to prevent invalid API tokens in the generation process.
\begin{tcolorbox}[breakable,width=\linewidth-2pt,boxrule=0pt,top=1pt, bottom=1pt, left=4pt,right=4pt, colback=gray!15,colframe=gray!15]
\textbf{Answer to RQ2:} Both components are essential for the performance of {\tool}. Removing hierarchical dependency and constrained decoding leads to a performance decrease. Among these, global dependency is the most important module, having the greatest impact on overall performance.
\end{tcolorbox}
\subsection{Efficiency of {\tool} (RQ3)} 

\noindent \textbf{Experimental Design.} To evaluate the efficiency, we measure the computational time for all approaches. Due to computational resource constraints, we use 20 concurrent processes for LLMs with up to 10B parameters and 10 concurrent processes for LLMs with over 30B parameters. For each approach, we specifically measure:

\begin{itemize}[leftmargin=*] 
\item \textbf{Base:} The time required for model inference with only incomplete function. 
\item \textbf{RAG:} The total time for retrieval and model inference. 
\item \textbf{De-Hallucination:} The total time for initial inference, retrieval, and a secondary inference. 
\item \textbf{{\tool}:} The total time for static analysis and model inference, reflecting its dependency analysis and decoding phases. \end{itemize}
To ensure reliable results, the reported computational time for each approach is averaged across all test samples. 

\noindent \textbf{Results.} As shown in Table~\ref{table:res_efficiency}, {\tool} achieves comparable efficiency to base models while greatly outperforming RAG and De-Hallucinator in computational time across all model sizes and benchmarks. 

\textbf{{\tool} maintains consistent efficiency across various benchmarks.} Different development scenarios greatly impact the inference time of the base model, as six LLMs achieve fast inference with an average time of 0.203s on {\dataset}-F and 0.256s on {\dataset}-M, respectively. RAG experiences a substantial increase in time, adding 0.457s on {\dataset}-F and 0.706s on {\dataset}-M due to the retrieval process. De-Hallucinator, which requires iterative generation, introduces additional delays, with an overhead of 0.474s on {\dataset}-F and 0.482s on {\dataset}-M. In contrast, {\tool} adds only 0.024s on {\dataset}-F and 0.02s on {\dataset}-M, maintaining minimal computational overhead. 
This benefit stems from the {\tool}'s lightweight design (i,e., not rely on the expensive retrieval or iterative processes.).

\textbf{{\tool} exhibits superior scalability with increasing model size.} For example, in the DeepSeekCoder family, the smallest model (DSC-1.3B) completes inference in an average of 0.211s across two benchmarks. {\tool} adds only 0.007s overhead, while RAG and De-Hallucinator add 0.576s and 0.333s, respectively. For the largest model (DSC-33B), despite the base inference time increasing to 0.314s, {\tool} maintains a low additional overhead of 0.017s. In comparison, RAG adds 0.652s, and De-Hallucinator adds 0.547s. These results demonstrate that {\tool} exhibits superior scalability, maintaining minimal overhead regardless of model size.
\begin{tcolorbox}[breakable,width=\linewidth-2pt,boxrule=0pt,top=2pt, bottom=2pt, left=4pt,right=4pt, colback=gray!15,colframe=gray!15]
\textbf{Answer to RQ3:} {\tool} achieves comparable efficiency to base models, adding only 0.022s overhead on average. {\tool}'s lightweight design and superior scalability make it an efficient framework for mitigating API hallucination across diverse development scenarios and model sizes.
\end{tcolorbox}

\subsection{Performance of {\tool} in Industrial Scenario (RQ4)}

\begin{table}
\centering
\caption{Evaluation results of baseline approaches and {\tool} in the industry scenario. The abbreviation of ``De-Hal.'' denotes ``De-Hallucinator''. \textbf{\raisebox{0.5ex}{\textasteriskcentered}} denotes statistically significant improvement (t-test with p-value < 0.001) over the baseline approaches.}
\label{table:res_industry}
\small
\begin{tabular}{c|ccccc}
\toprule
\textbf{Approach}                       & \textbf{EM(\%)} & \textbf{ES(\%)} & \textbf{MaHR(\%)} & \textbf{MiHN} & \textbf{Time(s)} \\ \midrule
\multicolumn{6}{c}{\cellcolor[HTML]{EFEFEF}PanguCoder-11B}                           \\ \midrule
\multicolumn{1}{c|}{Base}    &  13.63      &  50.53      &   85.45       &  1.06    &   0.205      \\ \midrule
\multicolumn{1}{c|}{RAG}     &  35.45      &   60.97     &       60.90     &   0.73    &  0.411      \\ \midrule
\multicolumn{1}{c|}{De-Hal.} &  13.76      &  48.72      &     86.23  &  1.11  &  0.479     \\ \midrule
\multicolumn{1}{c|}{{\textbf{\tool}}}     &  \textbf{60.90\raisebox{0.5ex}{\textasteriskcentered}}      &  \textbf{80.19\raisebox{0.5ex}{\textasteriskcentered}}      &   \textbf{25.45\raisebox{0.5ex}{\textasteriskcentered}}       &  \textbf{0.31\raisebox{0.5ex}{\textasteriskcentered}}    & \textbf{0.235}        \\ \midrule
\multicolumn{6}{c}{\cellcolor[HTML]{EFEFEF}PanguCoder-34B}                           \\ \midrule
\multicolumn{1}{c|}{Base}    & 13.63       &  48.27      &   84.54       &  1.05    &  0.206       \\ \midrule
\multicolumn{1}{c|}{RAG}     & 33.63       &  63.80      &   60.00     &  0.70    &    0.479     \\ \midrule
\multicolumn{1}{c|}{De-Hal.} &   13.76     & 49.80       &  85.32        &  1.04    & 0.501      \\ \midrule
\multicolumn{1}{c|}{{\textbf{\tool}}}     &  \textbf{58.18\raisebox{0.5ex}{\textasteriskcentered}}      &   \textbf{78.21\raisebox{0.5ex}{\textasteriskcentered}}     &   \textbf{23.63\raisebox{0.5ex}{\textasteriskcentered}}       & \textbf{0.30\raisebox{0.5ex}{\textasteriskcentered}}     & \textbf{0.239}        \\ \midrule
\end{tabular}
\end{table}

\noindent \textbf{Experimental Design.} To evaluate {\tool} in an industrial scenario, we construct a benchmark with 109 samples from Huawei's internal Java projects, following the process described in Section~\ref{sec:benchmark}. We use Huawei proprietary LLMs, PanguCoder-11B and PanguCoder-34B, as the base models for the experiments.

\noindent \textbf{Results.} The results in Table~\ref{table:res_industry} show that {\tool} outperforms baseline approaches in industrial scenarios, with statistically significant improvements (p-value < 0.001, Wilcoxon signed-rank test~\cite{wilcoxon1992individual}), while maintaining competitive inference efficiency.

\textbf{Effectiveness Analysis.} De-Hallucinator shows no improvement or even slight degradation compared to the base model. RAG achieves an EM of 35.45\%, but its high hallucination rate of 60.90\% on PanguCoder-11B limits its effectiveness. In contrast, {\tool} delivers significant improvements across all metrics. On PanguCoder-11B, it boosts EM by 71.79\%, ES by 31.52\%, and reduces MaHR by 58.21\% and MiHN by 57.53\% compared to RAG. These results demonstrate that merely providing relevant code snippets or iterative grounding is insufficient for industrial scenarios, while {\tool} effectively addresses these challenges by integrating project-specific dependencies and constrained decoding.


\textbf{Efficiency Analysis.} RAG and De-Hallucinator introduce substantial latency (e.g., 0.411s and 0.479s per sample on PanguCoder-11B) due to their reliance on retrieval and iterative processes. {\tool} achieves impressive efficiency, requiring only 0.235s per sample on PanguCoder-11B and 0.237s on PanguCoder-34B, adding just 0.030s and 0.033s overhead compared to base models. These results show that {\tool} is both effective and practical for industrial deployment, balancing performance with efficiency.
\begin{tcolorbox}[breakable,width=\linewidth-2pt,boxrule=0pt,top=2pt, bottom=2pt, left=4pt,right=4pt, colback=gray!15,colframe=gray!15]

\textbf{Answer to RQ4:} {\tool} demonstrates strong performance in industrial scenarios, achieving substantial improvements in both generation accuracy and hallucination mitigation while maintaining high efficiency.

\end{tcolorbox}

\section{DISCUSSION} \label{sec:discussion}
    \begin{figure*}[t]
\setlength{\abovecaptionskip}{-1pt}
    \centering
   \subfigure[Different project contexts provided in the input prompt]{
        \centering
        \includegraphics[scale=0.45]{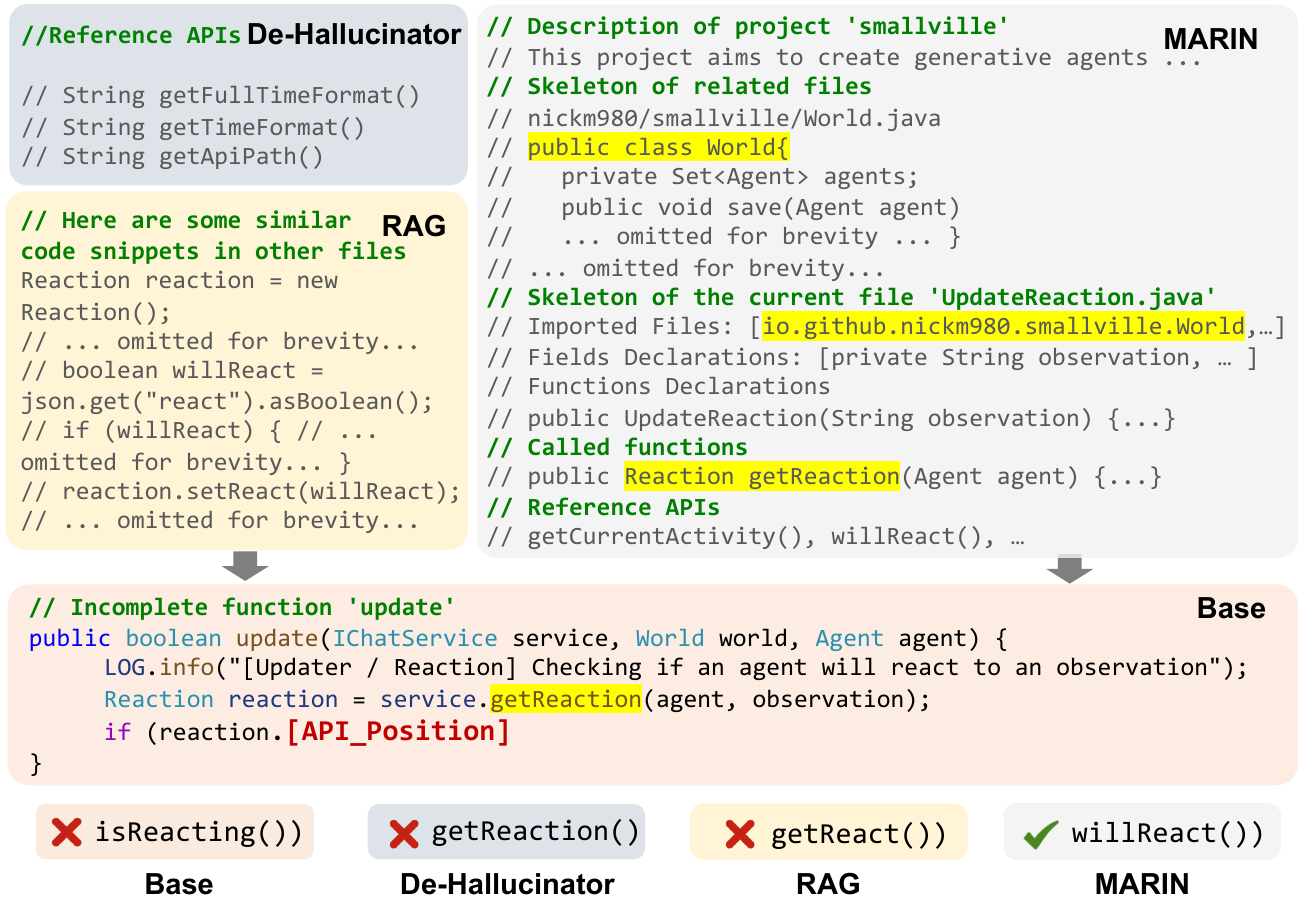} 
        \label{fig:case_1}
    }
    \subfigure[Generation process with Top-3 token logits]{
        \centering
        \includegraphics[scale=0.45]{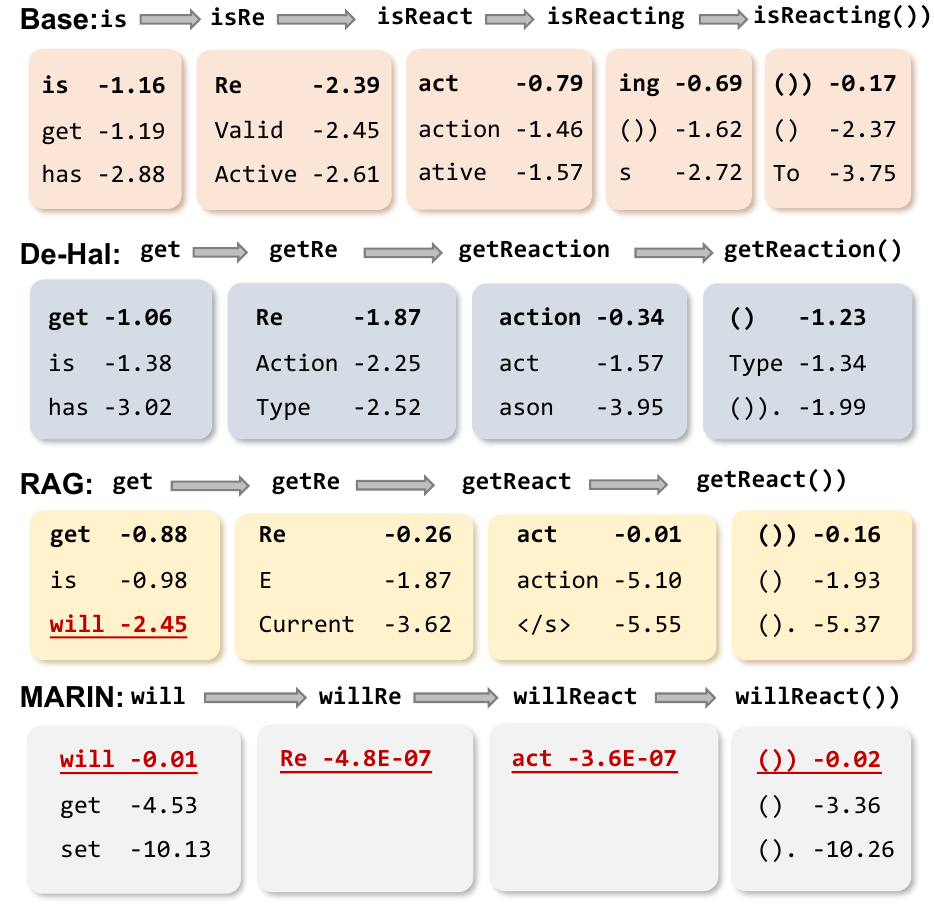}
        \label{fig:case_2}
    }
    \caption{Case study on API hallucination mitigation across Base, De-Hallucinator, RAG, and {\tool} using CodeLlama-7B.}
    \label{fig:case}
\end{figure*}

\subsection{Why does {\tool} work?}

The effectiveness of {\tool} lies in enriching the input context with structural project information and adaptively guiding the decoding process to adhere to valid APIs. As demonstrated in Figure~\ref{fig:case}, we revisit the example from the Introduction, where the {\tt update} function requires an API to determine an agent’s reaction to an observation. 

\subsubsection{Hierarchical dependency mining for enriching the input context with structural project information}

As shown in Figure~\ref{fig:case_1}, baseline approaches fail to provide sufficient context. The Base model relies solely on the incomplete function and generates the hallucinated API {\tt isReacting())}. De-Hallucinator retrieves relevant APIs but does not establish their connection to the function, producing the incorrect {\tt getReaction()}. Similarly, RAG retrieves similar code snippets but overlooks project-specific dependencies, leading to the invalid {\tt getReact())}. In contrast, {\tool} mines both local and global dependencies from the project such as the skeleton of {\tt Class World} and the implementation of {\tt getReaction} to create a structured and enriched input prompt. By providing a more comprehensive view of the project, {\tool} enables the model to accurately generate the API {\tt willReact())} while avoiding invalid alternatives.

\subsubsection{Dependency constrained decoding for adaptively guiding the generation process to valid API tokens.}

Figure~\ref{fig:case_2} shows the token logits for each approach, revealing differences in generation confidence. The Base model shows high uncertainty, with weak logits for its first token {\tt is} (-1.16). De-Hallucinator improves slightly, with stronger logits for {\tt get} (-1.06), and RAG retrieves broader context, resulting in logits of -0.88 for {\tt get} and -0.01 for {\tt act}. In contrast, {\tool} demonstrates near-certain logits for the correct token sequence, assigning -0.01 for {\tt will}, followed by deterministic logits for {\tt Re} (-4.8E-07) and {\tt act} (-3.6E-07). This precision is guided by the API prefix tree, ensuring that the generated tokens align with the project's valid APIs. Additionally, the final {\tt ())} token is accurately predicted through parameter validation, further enhancing the reliability of the generated code. By adaptively guiding the generation process, {\tool} ensures the generation complies with API specifications in the project, thus mitigating the API hallucination.

\subsection{Threats to Validity}

We identify four main threats to the validity of our study:

\textbf{Model Selection.} Although we have evaluated {\tool} on multiple models, including CodeLlama, DeepSeekCoder, and PanguCoder, many other LLMs remain unexplored, such as CodeQwen~\cite{codeqwen}, CodeGen~\cite{CodeGen}. However, the consistent improvements across current models suggest that our framework is highly generalizable. In the future, we will extend our evaluation to a wider range of models to further validate its scalability.

\textbf{Evaluation Tasks.} Our current implementation and evaluation are centered on Java projects. However, the design of {\tool} is language-agnostic, and we believe the framework can be extended to other programming languages with minimal adaptation, leveraging their respective dependency analysis.

\textbf{Prompt Design.} The effectiveness of {\tool} partially depends on our prompt template design. While we have conducted extensive experiments to validate our template, there might be room for further optimization. To mitigate this threat, we have open-sourced our prompt templates to facilitate community feedback and improvements. 

\textbf{Benchmark Curation.} Our benchmarks are created automatically. While automation enables scalability to handle extensive benchmarking tasks, it introduces potential concerns regarding dataset correctness and quality. In the future, we will enhance benchmark reliability by incorporating more extensive human validation or adopting hybrid dataset creation strategies.

\section{RELATED WORK}  \label{sec:related}
    \subsection{LLMs for Code Generation}

The impressive capabilities of LLMs in natural language processing have inspired researchers and companies to develop specialized models for code generation~\cite{CodeX,codellama,incoder,CodeGen,wavecoder,WizardCoder,mftcoder}. OpenAI proposes Codex~\cite{CodeX}, the earlier representative work of code LLMs. Subsequently, various models have emerged, such as Meta's InCoder~\cite{incoder} and CodeLlama~\cite{codellama}, Salesforce's CodeGen~\cite{CodeGen}, BigCode project's StarCoder series~\cite{starcoder,StarCoder2}, and Deepseek AI's DeepseekCoder~\cite{deepseek}. These code LLMs typically follow two development approaches: continuously training existing general models (e.g. CodeLlama), or training from scratch with code-specific data (e.g. DeepseekCoder). These advances have enabled commercial programming assistants like GitHub Copilot~\cite{copit} and Amazon CodeWhisperer~\cite{codewhisperer} demonstrate promising results in software development. Our proposed framework {\tool} can be applied seamlessly to current intelligent programming assistants to mitigate hallucinations in code generation, enhancing their accuracy and reliability.

\subsection{Hallucinations in LLMs}

While LLMs have shown remarkable performance in text generation, they frequently produce outputs that sound plausible but are incorrect, known as ``Hallucination'', severely impacting reliability and practical applications~\cite{text-1,text-2,relatedwork_hal1,relatedwork_hal2,relatedwork_hal3,relatedwork_hal4}. In the domain of code generation, hallucination poses similar challenges, particularly in scenarios involving API usage, where models may generate invalid API calls or misuse existing APIs. Liu~\textit{et al.}~\cite{liu2024exploring} conduct pioneering work in categorizing different types of code hallucinations. Zhang~\textit{et al.}~\cite{zhang2024llm} further explore the prevalence of code hallucinations in industrial settings, highlighting the unique challenges faced in real-world applications. Agarwal~\textit{et al.}~\cite{tian2024codehalu} propose a benchmark for detecting hallucinated code snippets on Python. To mitigate hallucination issues, Zhang~\textit{et al.}~\cite{zhang2024llm} retrieve project-specific code snippets to enhance generation. Eghbali \textit{et al.}~\cite{eghbali2024hallucinator} propose De-Hallucinator, which adds API references related to the model's predictions into the prompt. Jain \textit{et al.}~\cite{jain2024mitigating} introduce Documentation Augmented Generation (DAG), using API documentation from services like AWS and Azure to guide generation. Different from these approaches, our {\tool} combines project-specific structural context and dependency constrained decoding to ensure valid API generation, effectively mitigating hallucination in practical code generation scenarios.

\section{CONCLUSION} \label{sec:conclusion}
    In this paper, we propose
{\tool}, a framework designed to mitigate API hallucination based on hierarchical dependency analysis. By identifying local and global dependencies in the incomplete function and applying dependency constrained decoding, {\tool} ensures the generated APIs align with project requirements. Experiments show that {\tool} significantly improves API accuracy, reduces hallucination rates, and maintains efficiency with minimal overhead. In addition, {\tool}’s strong performance in industrial scenarios demonstrates its effectiveness in generating accurate and reliable APIs. As a solution to mitigate API hallucination in LLM-driven code generation, {\tool} contributes to the development of accurate and reliable code generation systems. In future work, we plan to extend {\tool} to a wider range of LLMs, additional programming languages, and optimize our prompt template.


\normalem
\balance
\bibliographystyle{ACM-Reference-Format}
\bibliography{ref}

\end{document}